\documentclass[submission,copyright,creativecommons]{eptcs}
\providecommand{\event}{LAFM 2013} 

\usepackage[utf8]{inputenc}
\usepackage[english]{babel}
\usepackage{xspace}
\usepackage{url}
\usepackage{bussproofs}
\EnableBpAbbreviations
\usepackage{paralist}

\usepackage{todonotes}
\font\ttfsyncopate syncopate at9pt
\font\ttflobster lobster at8pt

\font\ttfsyncopateST syncopate at13pt
\font\ttflobsterST lobster at12pt

\newcommand{\hetero}{{\ttfsyncopate Hetero}{\ttflobster Genius}\xspace}
\newcommand{\heteroST}{{\ttfsyncopateST Hetero}{\ttflobsterST Genius}\xspace}
\newcommand{\dynamite}{\texttt{Dynamite}\xspace}
\newcommand{\dynamiteThree}{\texttt{Dynamite3}\xspace}

\newcommand{\driver}{\emph{Analysis Driver}\xspace}

\newcommand{\manager}{\emph{Analysis Manager}\xspace}
\newcommand{\analaction}{\emph{analysis action}\xspace}
\newcommand{\analactions}{\emph{analysis actions}\xspace}

\newcommand{\pdocfa}{\textsf{PDOCFA}\xspace}
\font\ttfsyncopateT syncopate at14pt
\font\ttflobsterT lobster at12pt

\title{{\ttfsyncopateT Hetero}{\ttflobsterT Genius}: A Framework for Hybrid Analysis of Heterogeneous Software Specifications}
\author{
\qquad\qquad Manuel Giménez \qquad\qquad Mariano M. Moscato \qquad\qquad 
\institute{Departamento de Computación, \\
Facultad de Ciencias Exactas y Naturales,\\
Universidad de Buenos Aires}
\email{\{mgimenez, mmoscato\}@dc.uba.ar}
\and
Carlos G. Lopez Pombo 
\institute{Departamento de Computación, \\
Facultad de Ciencias Exactas y Naturales,\\
Universidad de Buenos Aires}
\institute{CONICET}
\email{clpombo@dc.uba.ar}
\thanks{The author would like to thank the MEALS project (EU FP7 programme, grant agreement No. 295261).}
\and
Marcelo F. Frias
\institute{Departmento de Ingeniería de Software,\\
Instituto Tecnológico de Buenos Aires}
\institute{CONICET}
\email{mfrias@itba.edu}
}
\def\titlerunning{\hetero: A Framework for Hybrid Analysis of Heterogeneous Software Specifications}
\def\authorrunning{Moscato, M.M., Lopez Pombo, C.G. et. al}

\begin{document}
\maketitle

\begin{abstract}
Nowadays, software artifacts are ubiquitous in our lives being an essential part of home appliances, cars, cell phones, and even in more critical activities like aeronautics and health sciences. In this context software failures may produce enormous losses, either economical or, in the worst case, in human lives. Software analysis is an area in software engineering concerned with the application of diverse techniques in order to prove the absence of errors in software pieces. In many cases different analysis techniques are applied by following specific methodological combinations that ensure better results. 
These interactions between tools are usually carried out at the user level and it is not supported by the tools.
In this work we present \hetero, a framework conceived to develop tools that allow users to perform hybrid analysis of heterogeneous software specifications.

\hetero was designed prioritising the possibility of adding new specification languages and analysis tools and enabling a synergic relation of the techniques under a graphical interface satisfying several well-known usability enhancement criteria. As a case-study we implemented the functionality of \dynamite on top of \hetero.

%
%
%
%
\end{abstract}


\section{Introduction}
Nowadays, software artifacts are ubiquitous in our lives being an essential part of home appliances, cars, cell phones, and even in more critical activities like aeronautics and health sciences. In this context software failures may produce enormous losses, either economical or, in the worst case, in human lives. Software analysis is an area in software engineering concerned with the application of diverse validation and verification techniques in order to prove the absence of errors in software pieces. 

Several languages and notations have been made available to help analysts and designers capture and model different aspects involved in software applications.  For instance, UML \cite{omg-sysml04,omg-ocl04} offers a range of diagrammatic notions, from class diagrams to state diagrams, collaboration diagrams, and so on.  This proliferation reflects the need to reduce the complexity of developing large systems as each language allows (teams of) engineers to address a specific view or phase of the development process. The same happens at the level of the formalisms that can formally support the use of such languages and methods. In summary, we face a scenario in which there is a multitude of modelling languages and supporting logics, and tools for processing such languages by reasoning in the underlying logics.
Thus, heterogeneity becomes a major source of complexity.
We find heterogeneity at the level of languages because different languages serve different purposes in software behaviour specification, and at the level of analysis as different techniques provide different results. 

The field of institutions \cite{goguen:cmwlp84} grew as an effort in providing formal foundations for software specification languages and analysis techniques. In \cite{meseguer:lc87}, Meseguer developed the categorical formalization of logical system by complementing the model theoretic view of a logic (institutions) with its deductive view (entailment system and proof calculus). In his work, Meseguer also introduced the notion of institution representations as a tool enabling reuse of proof systems, a limited view of heterogeneity. It was in \cite{tarlecki:sadt-rtdts95} where both institution morphisms and representations (also called co-morphisms) were extensively studied.
In \cite[Defs.~23~and~27, Sec.~4]{meseguer:lc87}, Meseguer extends the definitions of entailment and institution representation to work on theories.
In \cite[Prop.~5.2,~Thm.~5.3~and~Coro.~5.4, Sec.~5.1]{tarlecki:sadt-rtdts95}, Tarlecki proves general conditions under which a proof system for a richer logic\footnote{The word ``logic'' here is used as in \cite[Def.~6]{meseguer:lc87}.} $I'$ can be used to prove properties of specifications written in a poorer one $I$ provided there exists a map of logics from $I$ to $I'$. 
From now on we will call these operations \emph{$\rho$-translations}.\\


Many combinations of different tools have been depicted as methodologies for software analysis. Formal methods are usually divided into two categories: heavyweight and lightweight. These names refer to the amount of mathematical expertise needed during the process of proving a given property.
%
Modern software analysis methodologies departed long ago from the idea that heavyweight formal methods or lightweight ones are applied disregarding the relation between these tools.
We claim that enforcing these methodological directives as part of the process of software analysis produces better results. 

An example of this is \dynamite \cite{frias:tacas07}.
\dynamite is a theorem prover for Alloy \cite{jackson:acmtosem-11_2} in which the critical parts of the proof (carried out in a theorem prover implemented on top of the semi-automatic theorem prover PVS \cite{owre:cav96}) are assisted by the Alloy Analyzer with the aim of reducing both the workload and the error proneness introduced by the human interaction with the tool. Another use of model theoretic tools in relation to the use of theorem provers is the fact that they provide an efficient method for:
\begin{inparaenum}[\itshape a)\upshape]
\item the gain of confidence in the hypothesis brought into a proof,
\item the elimination of superfluous formulae appearing in a sequent,
\item the removal of minor modelling errors, and even 
\item the suggestion of potential witnesses for existential quantifiers.
\end{inparaenum}
All these actions are carried out by using the Alloy Analyzer in order to search for counterexamples for specific sets of conditions derived from the axioms in the specification and the property we want to prove. The result of this model-theoretical assistance for counterexample finding gives rise to a whole new class of analysis strategies resulting from a coordinated action of different tools over the same step of the proof.
These actions can not be understood neither as pure proof commands, nor as pure model-theoretical commands.
In such way, we call \emph{hybrid} those kinds of analysis that use a coordinated mixture of the both approaches. 


In this paper we will present the development of a general and scalable framework for building tools that allow the users to deal with hybrid analysis of heterogeneous specifications.
To test the capabilities of the \hetero framework we use it to build a new version of \dynamite.


\section{\heteroST: A framework for hybrid heterogeneous analysis}

Every analysis tool has to deal with four dimensions: the user interface, the conceptual analysis type, the languages and the analysis engine.
Most software out there are built to work with just one conceptual analysis type, provide support for only one language, and use just one specific analysis engine.
\hetero{} aims to be an extensible multi-language and multi-engine analysis tool, so its design must decouple the four dimensions as much as possible. 
It is worth noting there is a natural coupling between some of them: i.e., every engine is bounded with a specific language and the user interface is probably designed to work with one kind of conceptual analysis type.

\begin{figure}
\begin{minipage}[t]{0.475\textwidth}
\begin{center}
	\includegraphics[width=1\textwidth]{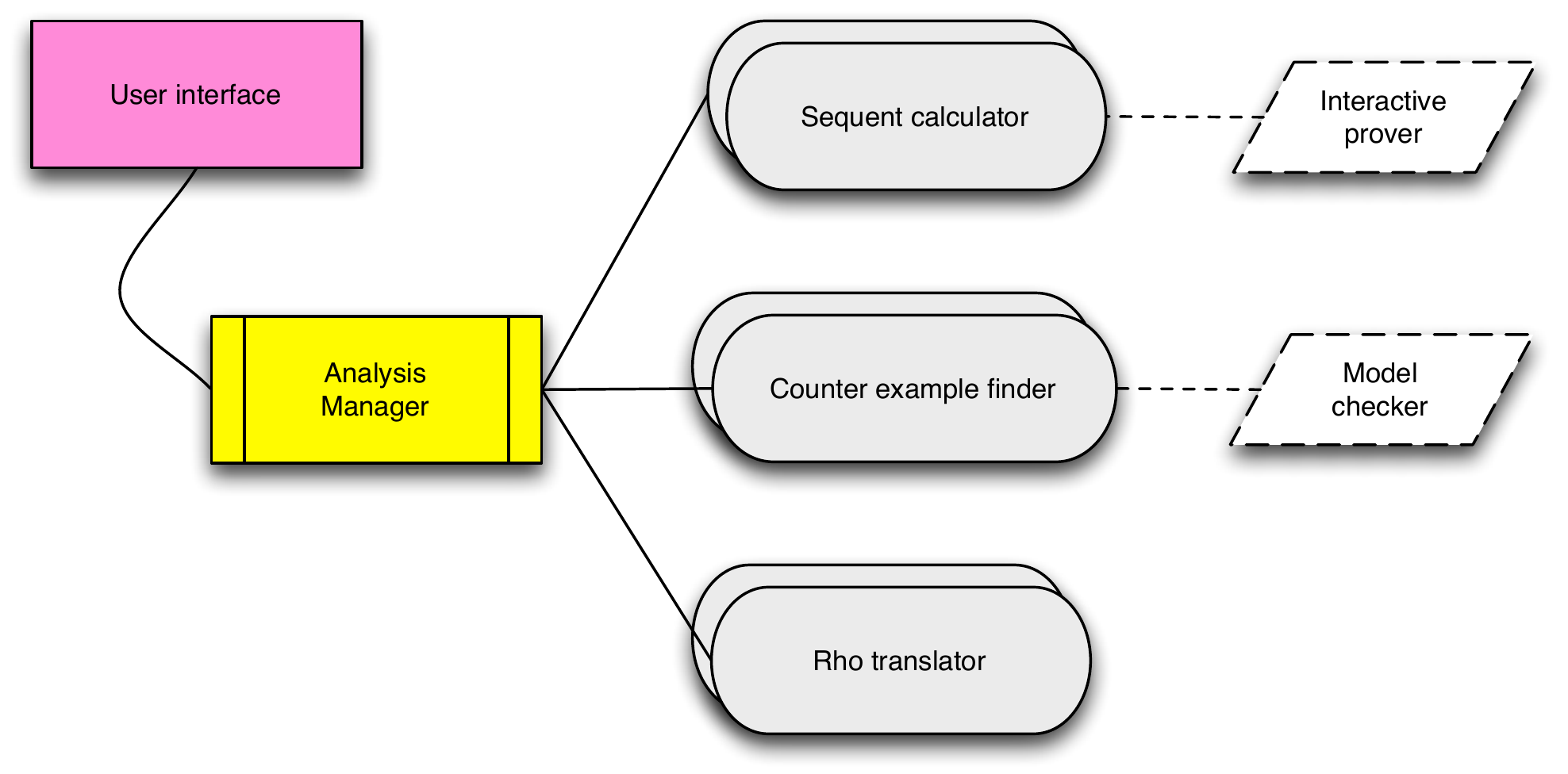}
	(\emph{a})
\end{center}
\end{minipage}\hspace{0.05\textwidth}
\begin{minipage}[t]{0.475\textwidth}
\begin{center}
	\includegraphics[width=1\textwidth]{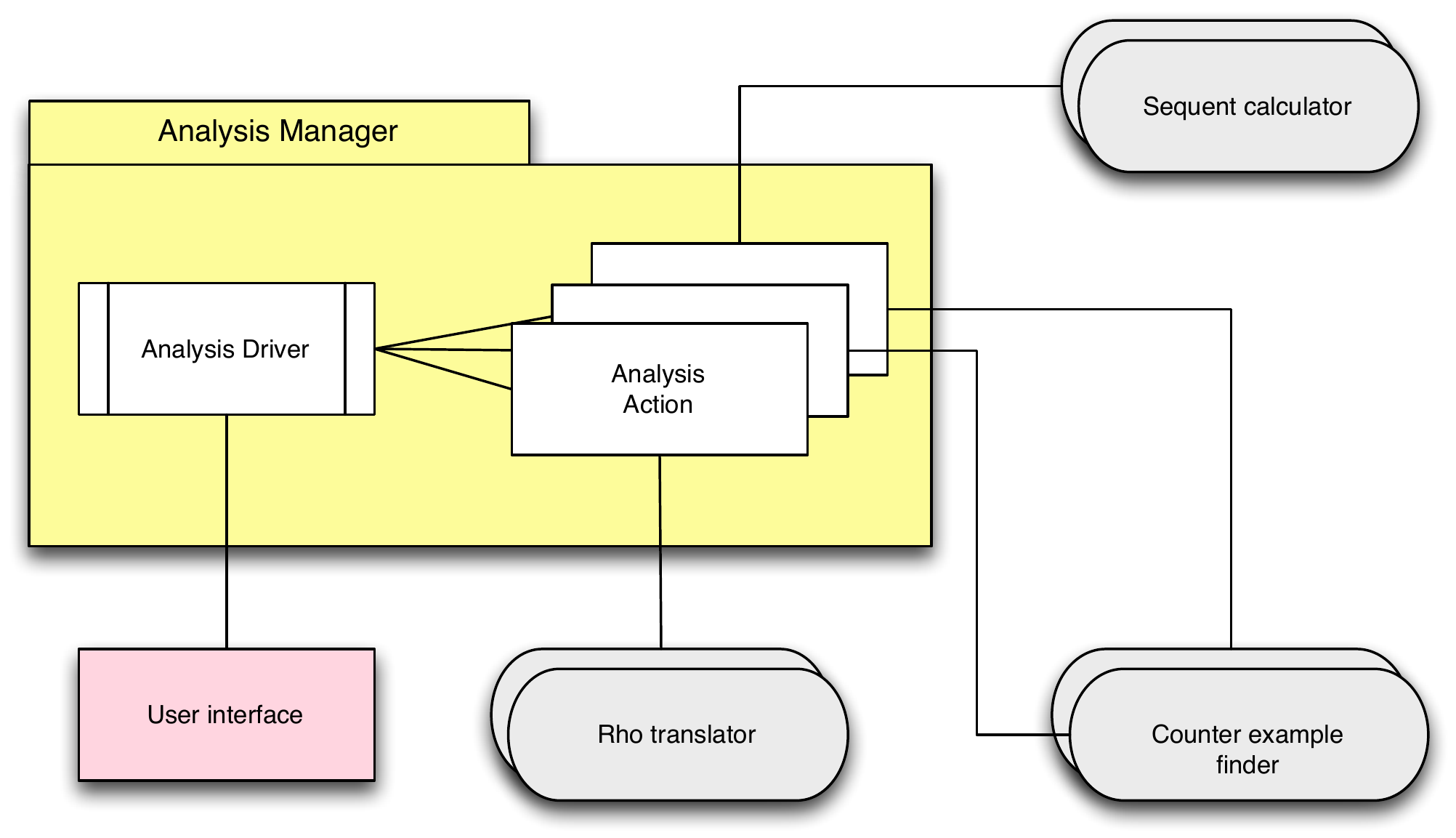}
	(\emph{b})
\end{center}
\end{minipage}
\caption{(\emph{a}) Global architecture of \hetero. \label{global} (\emph{b}) \manager architecture.\label{anal-manager}} 
\end{figure}

Figure \ref{global} shows the global architecture of \hetero{}.
The \manager keeps track of the current heterogeneous analysis, and drives its evolution.

\paragraph{Abstracting external engines}
It is well known that service abstraction is a very useful tactic to accomplish modifiability. We use it also to prevent our design to be tied to specific external analysis engines, introducing some components that abstract services offered by that external software.
We established three main families: sequent calculators, counter example finders, and $\rho$ translators.
Each family has a specific interface that captures and establishes the common behavior of its members.
Our design enables other families to be added easily without code modification as we will see in the next sections.

	
	A \textit{sequent calculator} is an entity that has just one responsibility: given a sequent and some rule of the sequent calculus, it must return the result of applying that rule over that sequent.
	Naturally most sequent calculus interactive provers can be used as backends for this type of component. It is worth noting that each concrete calculator has its own set of rules.
	The intra language rules of the calculus actually implemented by \hetero{} is going to be limited by the sum of all rules provided by the concrete sequent calculators.
	A \textit{counter example finder} is a component that given a formula tries to answer whether or not a counter example exists for that formula.	
	A \textit{$\rho$ translator} offers language translation services by translating sequents and specifications between languages.
	None of the mentioned components need to know anything about the current analysis, accomplishing in this way the desired decoupling between the engines and the analysis management.



Note that the same external engine can be used as backend for more than one concrete component.
Our architecture does not force a specific way to communicate with external applications; each component is free but also responsible to choose the method it considers appropriate.


\paragraph{Analysis Manager}


The following are actions that the user might want to perform over the current analysis: apply certain sequent calculus rule over some analysis node, validate some sequent by searching for a counterexample, prune the analysis tree, change languages at some point of the analysis, etc.
It is clear that the specific steps to perform each of the mentioned actions are very different: to apply the rule, a sequent calculator must be used and several new nodes may have to be added to the tree; on the other hand when validating the sequent no nodes will ever be added.

We introduced the idea of \analaction{} to model any kind of action the user might want to perform over the analysis. Each \analaction{} is responsible for knowing and performing all the specific steps to actually apply the action it models.
One \analaction{} may interact with none, one, or several of the external engine abstractions mentioned in the previous section.

As shown in figure \ref{anal-manager}, the \driver{} does not need to interact directly with any of them, thanks to the \analactions, increasing the semantic coherence of our design.
This component is responsible for keeping the current state of the analysis tree, but as we explained, the changes over it are made by \analactions{}. The analysis tree design enables other languages to be added easily without code modification.




\paragraph{User interface}



The user interface of a software like \hetero{} must present a clear outlook of the current analysis and provide an easy and intuitive way of interacting with it. According to \cite{beckertevaluating2012}, the way in which the current status of the proof is shown (formulae, sequents) is crucial for the usability. Our efforts were focused on achieving these objectives.

Visualization of tree structures is a research area on its own, so we decided to develop the interface of \hetero{} using a mature framework: JUNG2\footnote{\url{http://jung.sourceforge.net}}.
The Java Universal Network/Graph Framework is a software library that provides a common and extendible language for the modeling, analysis, and visualization of data that can be represented as a graph or network. It is written in Java and it is open source.

In \hetero{} almost all interactions are done by \textit{point \& click}: just left click any analysis node, and a contextual menu will offer all the \analactions{} applicable over the clicked node. In case that the \analaction{} being applied needs a formula from the user, she or he must use the keyboard to provide it.


\section{\dynamiteThree: Implementing \dynamite on top of \heteroST}

In order to implement a new version of \dynamite using the \hetero framework, we needed to provide a \emph{sequent calculator} for Alloy.
As mentioned above, the previous version of \dynamite \cite{moscato:ictac10} is an interactive theorem prover for Alloy, based on sequent calculus.
Then, we only had to wrap it with a new component that plays the role of an abstraction layer.
The formal background of \dynamite can be found in \cite{frias:tacas07} and \cite{frias:icfem04}, where it is proved that there exists a semantic preserving translation of the Alloy specification language to theories in an extension of fork algebras \cite{frias02}\footnote{The interested reader will find model-theoretic flavoured proofs based on the semantics of the languages in that work, but they can be easily restated within the framework of general logics by just bringing into the definitions of the translations their action on morphisms.}.
The class of algebraic structures considered for interpreting Alloy is the class of \emph{point-dense omega closure fork algebras} (\pdocfa). 
Resorting to PVS's higher-order logic we constructed a semantics embedding of \pdocfa.
Then, we could be able to use the PVS prover to provide a theorem prover for Alloy.


To test the \hetero ability of supporting heterogeneous proofs, we also implemented a concrete \emph{sequent calculator} for \pdocfa.
This calculator was developed by reusing much of the code of \dynamite, since the embedding of the language of \pdocfa in PVS was already implemented in that tool.

We also implemented an Alloy \textit{counter example finder} using the Alloy Analyzer as backend.
Finally we developed an Alloy to \pdocfa $\rho$ translator, which internally relies on the same translation used by the Alloy sequent calculator mentioned above.

Several new \analactions were added: one for each sequent calculus rule provided by our \textit{sequent calculators}, one to use the Alloy \textit{counter example finder} to look for counter examples of the sequent of a given analysis node, one to switch languages during the analysis using the $\rho$ translator, and a few more to manipulate the analysis tree (like pruning a subtree).
We also added some \analactions{} that make use of interactions between different engines in order to provide the specific \dynamite commands, such as validated case or pruning of goals \cite{moscato:ictac10}.

\dynamiteThree is available at \url{http://www.dc.uba.ar/dynamite/heterogenius}.


\section{Conclusions, related and further work}

In the line of the tools for heterogeneous analysis one remarkable piece of work is Hets.
According to \cite{mossakowski:tacas07}, Hets is intended for use as a proof manager of complex systems, specified using different formal languages.
But once a specific obligation has to be proved, the proving process must be carried out using a specific external tool.
Hets serves as an organizer of information related with which goals have been proved, which ones have not and how elements specified using different languages are related with each other.
Instead, \hetero is much more involved in the process of actually proving the goals.
As explained above, it manages each proof of the specification, allowing the use of different external tools even when proving a single goal.
Even more, \hetero allows the interaction between different external tools and thus gives the user the possibility of performing hybrid analysis on the specification under study.
The way in which each element is related with another one specified in a different language, is established through the $\rho$ translators mentioned in previous sections.
The management of the dependencies of the obligations being proved is one of the next steps in the development of \hetero.


In this work we showed some of the relevant theoretical and practical issues behind the implementation of tool support for hybrid analysis of heterogeneous software specifications.
Theoretic results were put in second place to leave enough space for the intuitions and the discussion on the development decisions.
Finally we showed how the architecture of \hetero enabled the reengineering of \dynamite, a heterogeneous analysis tool.

The reader should notice that even when the implementation of \dynamite requires certain level of heterogeneity in the language supporting the analysis, it fails in exemplifying how \hetero can help in analyzing a software artifact described by the interactions of components described in different logical languages.
Regarding this we are working on the implementation of an heterogeneous specification language presented in \cite{lopezpombo:phdthesis}.

\bibliographystyle{eptcs}
\bibliography{lafm2013,bibdatabase}

\end{document}